\documentclass[a4paper,12pt]{article}
\usepackage[a4paper, left=1in,right=1in,top=1in,bottom=1in]{geometry}
\usepackage{soul}
\usepackage[usenames,dvipsnames]{color}
\usepackage{amsmath, amssymb, slashed, epsf, color, graphicx, latexsym}
\usepackage{epsfig}
\usepackage{graphics}
\usepackage{float}

\usepackage{jheppub}

\newcommand{\be}{\begin{equation}}
\newcommand{\ee}{\end{equation}}
\newcommand{\bea}{\begin{eqnarray}}
\newcommand{\eea}{\end{eqnarray}}
\def\bse{\begin{subequations}}
\def\ese{\end{subequations}}
\newcommand{\IR}{\mathbb{R}} \newcommand{\IC}{\mathbb{C}}
 
\def\IZ{\relax\ifmmode\hbox{Z\kern-.4em Z}\else{Z\kern-.4em Z}\fi}
\newcommand{\IS}{\mathbb{S}} 

\newcommand{\non}{\nonumber \\}

\def\half{\frac{1}{2}}

\def\al{\alpha} 
  \def\eps{\epsilon}
\def\sig{\sigma} \def\bsig{{\bar \sigma}} \def\hsig{{\hat \sigma}} \def\tsig{{\tilde \sigma}}

\def\calE{\mathcal{E}} 
\def\div{{\rm div}}	
\def\rf{{\rm ref}} \def\rfp{{\rm ref'}} \def\br{{\rm bare}}
\def\bV{{\bar V}}

\def\presub{\vspace{.5cm} \noindent}

\def\bi{\begin{itemize}} \def\ei{\end{itemize}}

\def\({\left(} \def\){\right)}
\def\[{\left[} \def\]{\right]}
\def\<{\left<} \def\>{\right>}

\title{Regularized phase-space volume for the three-body problem}
\author[a]{Yogesh Dandekar,}
\author[a]{Barak Kol,}
\author[a]{Lior Lederer,}
\author[b]{Subhajit Mazumdar }
\affiliation[a]{Racah Institute of Physics, Hebrew University, Jerusalem 91904, Israel} 
\affiliation[b]{Center for Theoretical Physics, Seoul National University, Seoul 08826, Korea}
\emailAdd{yogesh.dandekar, barak.kol@mail.huji.ac.il,\\ lior.lederer.40@gmail.com, subhajitmazumdar@snu.ac.kr} 

\abstract{The micro-canonical phase-space volume for the three-body problem is an elementary quantity of intrinsic interest, and within the flux-based statistical theory, it sets the scale of the disintegration time. While the bare phase-volume diverges, we show that a regularized version can be defined by subtracting a reference phase-volume, which is associated with hierarchical configurations. The reference quantity, also known as a counter-term, can be chosen from a 1-parameter class. 

The regularized phase-volume of a given (negative) total energy, $\bsig(E)$, is evaluated. First, it is reduced to a function of the masses only, which is sensitive to the choice of a regularization scheme only through an additive constant. Then, analytic integration is used to reduce the integration to a sphere, known as shape sphere. Finally, the remaining integral is evaluated numerically, and presented by a contour plot in parameter space. Regularized phase-volumes are presented for both the planar three-body system and the full 3d system. In the test mass limit, the regularized phase-volume is found to become negative, thereby signalling the breakdown of the non-hierarchical statistical theory.

This work opens the road to the evaluation of $\bsig(E,L)$, where $L$ is the total angular momentum, and it turn, to comparison with simulation determined disintegration times.}

\begin{document}

\maketitle

\section{Introduction}

The three-body problem involves three point-like objects moving under the influence of their mutual Newtonian gravitational attractions. It was defined and studied already by Newton (1687) in the context of the motion of the moon \cite{Principia}, following his success in solving the two-body problem. Over the centuries, its study triggered and influenced whole new fields, including perturbation theory, chaos and topology, which is a testament to its richness. Yet, it remains arguably the oldest open problem in physics, see the reviews \cite{Valt_Kart_book_2006,Valtonen_etal_book_2016}.

Within this system, Poincar\'e (1890) discovered the phenomenon of chaos  \cite{Poincare_1890}, namely extreme sensitivity to initial conditions, and since then the system is believed not to have a deterministic solution in closed-form. Agekyan and Anosova (1967) \cite{Agekyan_Anosova_1967} began its exploration through computer simulations, see also references within \cite{Valt_Kart_book_2006}. Monaghan (1976) \cite{Monaghan_1976} presented a statistical theory towards a statistical solution, and more recently, \cite{Stone_Leigh_2019} presented a statistical prediction in closed-form, based on that theory, and \cite{Ginat_Perets_2020} presented an improvement that incorporates intermediate hierarchical phases. 

However, all of these statistical approaches rely on the introduction of a parameter that is not part of the problem's data, namely, the strong interaction radius, and accordingly all the ensuing statistical predictions depend on it. To remedy that, \cite{flux_based} re-examined the basis for the statistical theory, and presented a different statistical theory, one which relies on the flux of phase-space volume, rather than the phase-space volume itself. This flux-based approach eliminates the strong interaction radius. Together with the emissivity-blindness assumption, it provides a rather simple expression for the probability for each of the bodies to escape. This prediction displays a leap in the fit of the statistical theory with computer simulations, to about 1\%, thereby becoming the most precise statistical theory to-date for the three-body problem \cite{MTL_2020,simulate}. 

Within the flux-based theory, the differential decay time $d\Gamma$ is predicted to be given by \be
	d\Gamma(u) = \frac{1}{\bsig_\chi}\, \calE(u)\, dF(u)
	\label{bsig_in_flux_based}
\ee
where $u$ is a collective notation for all outcome parameters that the decay rate is distributed over, $\bsig_\chi$ is the regularized chaotic phase-volume (short for phase-space volume) of the system, $\calE(u)$ is the chaotic emissivity function (equivalently, absorptivity) and $dF(u)$ is the distribution of asymptotic phase-volume flux. The theory provides a closed-form expression for $dF(u)$, while $\calE(u)$ requires an independent measurement through simpler simulations and/or an analytical model.  In this sense, the flux-based theory provides an exact reduction of the statistical solution to the three-body problem. This relation is motivated by an analogy with a particle moving inside a leaky container, where $\bsig_\chi$ is analogous to the container volume. Another way in which the flux-based theory goes beyond previous statistical approaches is in modelling not only the outcome statistics, but also the distribution of decay times. 

In this paper, we focus on the regularized phase-volume. It is a $u$-independent normalization, which sets the scale for the decay rate. $\bsig_\chi=\bsig_\chi(E,L)$ denotes the regularized chaotic phase-volume given the conserved charges, the total energy $E$ and the total angular momentum $L$. The chaotic phase-volume is heuristically defined as the part of phase space that displays chaotic motion. The goal of this paper is to determine a related quantity, the regularized phase-volume $\bsig(E)$.  $\bsig(E)$ is easier to determine than $\bsig_\chi(E,L)$, yet it still requires to fully define and evaluate the regularization. We leave to future work the determination of other variants of the regularized phase-volume that would enable comparison with measured decay rates. 

This paper is organized as follows. We start in section \ref{bss} by defining the phase space volume, followed by the definition of the proposed regularization scheme. The integral that defines the regularization is evaluated in section \ref{sec:eval}: first, through analytic integration, to the extent possible,  and the remainder through numerical integration. The regularization process is repeated for the planar three-body problem in \ref{sec:planar}. Finally, a summary and discussion are presented in section \ref{sec:disc}.

\section{Definition of regularization}
\label{bss}

\subsection{Setup, reduction and rate of divergence}

\noindent {\bf Setup}. The Newtonian three-body problem can be defined through the Hamiltonian \be
 	H = T+V = \sum_{a=1}^{3} \frac{p_a^2}{2m_a} - \sum_{a<b} \frac{\al_{ab}}{r_{ab}} ~,
\label{def:H}
\ee
which is a function of the three positions $\vec{r}_a, ~a=1,2,3$ and the three momenta $\vec{p}_a$, which together define the system's dynamical variables; $H$ is also a function of the three mass parameters $m_a$; and finally, we define the relative position vectors $\vec{r}_{ab} := \vec{r}_a - \vec{r}_b$ and the potential strength constants $\alpha_1 \equiv \alpha_{23} : = G\, m_2\, m_3$ and similarly by cyclic permutations.
 
Given the conserved charges, the total energy $E$, and the total angular momentum $\vec{L}$, the system's phase space volume is defined by \be
	\sigma(E,\vec{L})  := \int \( \prod_{a=1}^{3}\, d^3 r_a\, d^3p_a \)  \delta^{(3)}(\vec P_{CM}) ~ \delta^{(3)}(\vec R_{CM}) ~ \delta(H-E) ~ \delta^{(3)}(\vec J- \vec L) ~,
\label{def:sigEL}
\ee
	where the center of mass position $\vec{R}_{CM}$, the total (center of mass) momentum $\vec{P}_{CM}$ and the total angular momentum $\vec{J}$ are given by the usual definitions $\vec{R}_{CM} := \frac{1}{M} \sum_{a=1}^{3}\, m_a\, \vec{r}_a$, $\vec{P}_{CM} := \sum_{a=1}^{3}\, \vec{p}_a $, and $\vec J := \sum_{a=1}^{3} \vec r_a \times \vec p_a$, where $M:=\sum_{a=1}^3\, m_a$ is the total mass.

 In this paper, we are interested in a simpler quantity, the phase-volume for a given total energy, but ignoring the total angular momentum, namely \be
	\sigma(E)  := \int \( \prod_{a=1}^{3}\, d^3 r_a\, d^3p_a \)  \delta^{(3)}(\vec P_{CM}) ~ \delta^{(3)}(\vec R_{CM}) ~ \delta(H-E) ~.\label{def:sigE}
\ee
The only difference with respect to \eqref{def:sigEL} is the omission of $\delta^{(3)}(\vec J- \vec L)$. $\sigma(E)$ is related to $\sigma(E,\vec{L})$ through $\sigma(E) = \int d^3L ~\sigma(E,\vec L)$. In addition to the dependence on $E$, $\sigma(E)$ depends also on the problem's parameters, namely $ \sigma(E) \equiv \sigma(E;\{m_a\}_{a=1}^3)$. 

\presub {\it Rationale behind focusing on $\bsig(E)$}. Comparison with measured decay rates requires the regularized chaotic phase-volume $\bsig_\chi(E,L)$ \eqref{bsig_in_flux_based}. A major challenge on the road to its evaluation is to fully define the quantity and its regularization. The evaluation of $\bsig(E)$ is simpler than that of $\bsig_\chi(E,L)$, while $\bsig(E)$ still requires to overcome this central challenge.  Hence, it is a worthwhile goal for the present work.

After a regularization method is available, one could further refine it into a definition of $ \bsig(E,L)$ and proceed to evaluation. Next, the evaluation of $\bsig_\chi(E,L)$ would require to separate out regions in phase space that describe fully regular time evolutions. However, experience with simulations shows that chaotic time evolutions are much longer than regular ones. Hence, the chaotic component of $\bsig(E,L)$ is expected to dominate it, namely $\bsig_\chi(E,L) \simeq \bsig (E,L)$.

The expected connection between $\bsig(E)$ and simulation results can be summarized by  \be
	\bsig(E) ~~ \to ~~ \bsig(E,L) \simeq \bsig_\chi(E,L) ~~ \to ~~   \mbox{comparison w. measurement} ~. 
\ee

\presub {\bf Reduction}. The momentum integration in the definition of $\sigma(E)$ can be performed using the method of Sec. 3.4 of \cite{flux_based}, where one introduces auxiliary integration variables conjugate to $\vec{P}_{CM}, E$, to obtain \bea 
	\int \( \prod_{a=1}^{3}\,  d^3p_a \)  \delta^{(3)}(\vec P_{CM}) ~  \delta(H-E) &=&   \mbox{Vol}(\IS^5) \, \left(\frac{M_3}{M}\right)^{3/2} \, (2(E-V))_+^2 \non
						&=&  4 \pi^3\, \left(\frac{M_3}{M}\right)^{3/2} \, (E-V)_+^2
\label{momenta_integ}						
\eea
 where we have defined the elementary symmetric functions of the masses \bea
 M &:=& m_1 + m_2 + m_3 \non
 M_2 &:=& m_2 \, m_3  + m_3 \, m_1  + m_1 \, m_2 \non
 M_3 &:=& m_1\, m_2\, m_3 ~,
\eea 
and where the ramp function is defined by \be
 (x)_+ : = \begin{cases}
 x 	& 0 \le x \\
 0 	& x \le 0
\end{cases} ~\equiv~  \frac{x + |x|}{2} \equiv x\, \Theta(x)
\label{def:ramp}
\ee 
and $\Theta$ is the Heavyside $\Theta$-function. 

The dependence of $\sigma(E)$ on $E$ can be factored out  as follows. Rewriting $(E-V)_+^2 = |E|^2\, (|V/E|-1)_+^2$, where we assume $E<0$, one notices that the integrand that defines $\sigma(E)$ depends only on the rescaled position variables $\vec{\tilde{r}}_a = |E|\, \vec{r}_a$. Transforming to these variables, and then omitting the tildes \footnote{Whenever the reader is unsure about whether we are using the (original) position variables $(r)$ or the rescaled position variables $(\tilde r)$, the appropriate factors of Energy $(|E|)$ can be reinstated by dimensional analysis.} $\tilde{r} \to r$, one finds \be
	\sigma\(E;\{m_a, \al_a\}\) = 4 \pi^3\, \left(\frac{M_3}{M}\right)^{3/2} \, \frac{1}{|E|^4} \; \sigma\( \{\al_a\} \) ~,
\label{reduction}
\ee
 where $\sigma\( \{\al_a\} \)$  is defined by \be
	\sigma\( \{\al_a\} \) :=  \int \(  \prod_{a=1}^{3} d^3 r_a \) ~\delta^{(3)}(\vec{R}_{CM}) \[ \( \sum_{a<b} \frac{\alpha_{ab}}{r_{ab}} \)- 1\]_+^2 ~.
\label{def:red_sig}
\ee

The reduction \eqref{reduction} applies more generally to three-body problems with $1/r$ potentials whose potential strength constants, $\alpha_a$, are arbitrary, rather than being fixed by the masses. In this context, the dependence on the masses and the energy is separated, and it remains to determine the dependence on $\alpha_a$.

\presub {\bf Divergence}. The phase-volume  \eqref{def:sigE} diverges. This can be seen by considering the reduced phase-volume \eqref{def:red_sig}. When checking for divergence, one should consider both the asymptotic regions in the integration domain, and locations where the integrand diverges. The first asymptotic region to consider is the case where all three bodies are far apart. In this case, the potential is close to $0$, hence the ramp function $(E-V)_+$ vanishes, and together with it, the whole integrand.

\begin{figure}
\centering \noindent
\includegraphics[width=14cm]{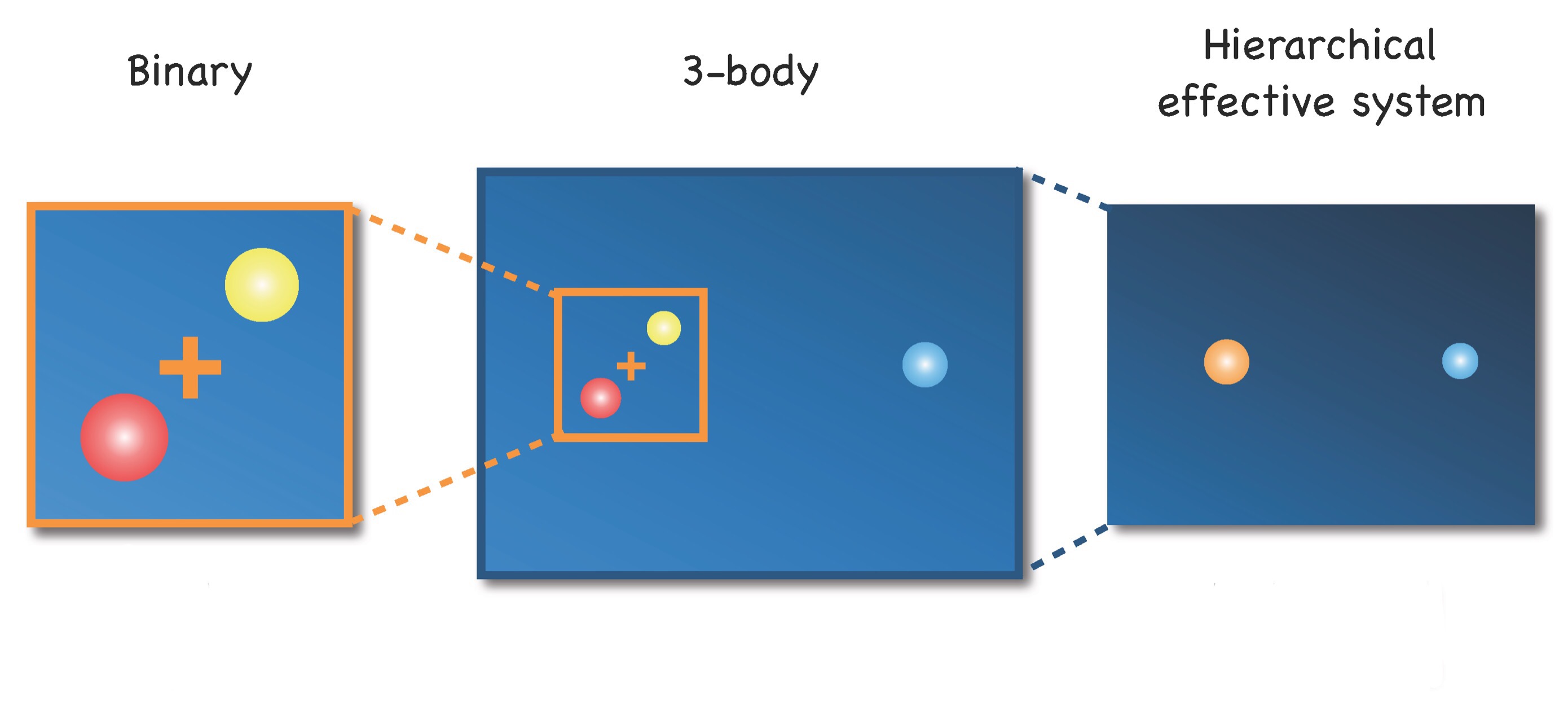} 
\caption[]{Hierarchical configurations lead to a divergence of the phase space volume. Middle: A hierarchical configuration. Left: the binary subsystem. Right: the hierarchical effective system, the rightmost body is the escaper.}
 \label{fig1:coord_systems} \end{figure}

The second asymptotic region to consider is that of hierarchical configurations, where two of the bodies are separated by less than a certain critical distance, while the third body is arbitrarily far away, see figure \ref{fig1:coord_systems}. In this case, $V$ is dominated by the binary potential and is almost independent of the relative separation. Hence, the integrand is almost constant, while the integration domain is infinite, and therefore the integral diverges. We shall look at the rate of divergence momentarily.

Finally, the integrand in \eqref{def:red_sig} diverges whenever two of the bodies approach each other. Denoting the binary separation by $r_B$ , one has $V \sim -1/r_B$ as $r_B \to 0$, and hence the integrand $\sim (E-V)+^2 \sim 1/r_B^2$. However, the measure contributes $d^3 r_B \sim r_B^2\, dr_B$ which renders the integral convergent at the neighborhood of this point.

\presub {\it Rate of divergence}. In hierarchical configurations, the three-body potential, defined in \eqref{def:H}, can be approximated by the effective potential 
 \be
V_{F,s} := -\frac{\alpha_B}{r_B}   -\frac{\alpha_F}{r_F}
\label{def:V_F}
\ee
where $s=1,2,3$ is the identity of the escaper (the widely separated tertiary), the binary potential strength is $\alpha_B:= G m_a m_b$ where $m_a, m_b$ are the binary components, the binary separation vector is given by $\vec{r}_B:= \vec{r}_a - \vec{r}_b $, and similarly the effective potential strength is given by $\alpha_F:= G m_s (m_a +m_b)$ and $\vec{r}_F= \vec{r}_s - ( m_a \vec{r}_a + m_b \vec{r}_b)/(m_a+m_b)$. $V_{F,s}$ describes a pair of decoupled Keplerian two-body problems. Note that the small term $(-\alpha_F/r_F)$ must be included in $V_{F,s}$ in order to account for the two-body Keplerian divergence in phase-volume due to the relative, or effective, motion. This phase-volume also has the interpretation of the Keplerian delay time, or the quantum scattering phase.

In the effective variables, the measure becomes \be
 \int \( \prod_{a=1}^{3} d^3 r_a \) ~\delta^{(3)}(\vec{R}_{CM}) = \int d^3r_F\, d^3 r_B ~.
\ee
Let us denote the rate of divergence of a quantity $X$ by $\div(X)$, whose formal meaning is that $X-\div (X)$ is finite. Of course, $\div (X)$ is defined only up to finite terms. Now, the preceding discussion can be summarized by \be
\div \( \sigma\( \{\alpha_a\} \) \) = \sum_{s=1}^3 \int^\infty d^3 r_{F,s} \int d^3 r_{B,s}  \( - V_{F,s}-1 \)_+^2 ~,
\label{div_sig1}
\ee
where the integration limit for $d^3r_{F,s}$ stresses the divergent neighborhood.

Recall that the integration domain in \eqref{div_sig1} is over rescaled position variables,  defined above \eqref{reduction}, and these have the dimensions of potential strength constant, namely $\[ \tilde{r} \] = \[ r\, E \] = \[ \alpha \]$. In fact, $\alpha_B$ is the only quantity in the binary system with these dimensions, and hence the $d^3 r_B$ integral is proportional to $\alpha_B^3$. Similarly, the $d^3 r_F$ integral must be proportional to $\alpha_F^3$ times a dimensionless divergent factor. Altogether, \be
\div \( \sigma\( \{\alpha_a\} \) \) \propto \sigma_0 := \sum_{s=1}^3 \( \alpha_{B,s} \, \alpha_{F,s} \)^3 = (\alpha_1 + \alpha_2)^3 \alpha_3^3 + cyc. 
\label{div_sig2}
\ee
More precisely, the rate of divergence can depend on three independent cut-offs, one for each $s=1,2,3$. 

Finally, the rate of divergence can be expressed in full by changing to the following dimensionless variables \be
u_B := \frac{\alpha_B}{\tilde{r}_B} \equiv \frac{\alpha_B}{|E| \, r_B} \qquad u_F := \frac{\alpha_F}{|E| \, r_F} ~.
\label{def:u}
\ee
in terms of which the rate of divergence becomes \be
\div \( \sigma\( \{\alpha_a\} \) \) = \sigma_0  \int_0 \frac{d^3 u_F}{u_F^6}  \int \frac{d^3 u_B}{u_B^6}  \( u_B + u_F -1 \)_+^2 ~
\label{div_sig3}
\ee
Again, the explicit integration limit for $u_F$ denotes the location of the divergence.

\subsection{Regularization}\label{regsec}

While the phase-volume defined by \eqref{def:sigE}  is divergent, a meaningful finite part $\bsig$ can be extracted from it by a process of regularization, whereby a divergent part is subtracted as follows \be 
	\bsig(E) := \sigma(E) -  \sigma_{\rf}(E) \qquad \mbox{at integrand level,}
\ee
 where $\bsig$ and $\sig$  are known as the regularized and the bare parts, respectively, and $\sig_{\rf}$ is known as the reference part or the counter-term. The subtraction should be performed at the level of the integrands, thereby avoiding the appearance of infinite quantities. We note that whereas the bare value is positive by definition, the regularized value, being a difference, could have either sign. 

We define the reference term by  \be
\sigma_{\rf}(E) := \sum_{s=1}^3 \int_{D_s} \( \prod_{a=1}^{3}\, d^3 r_a\, d^3p_a \)  \delta^{(3)}(\vec P_{CM}) ~ \delta^{(3)}(\vec R_{CM}) ~ \delta(H_{F,s}-E) 
\label{def:sig_ref}
\ee
 where the first difference with respect to \eqref{def:sigE} is the replacement of the full potential $V$, by the effective potential $V_{F,s}$ given in \eqref{def:V_F}, namely $H_{F,s}:= T + V_{F,s}$,  where $T$ is still given by \eqref{def:H}. The second and last difference is the specification of the integration domains $D_s$, which are defined to be all points in phase space such that \be 
	E_B  \le E/2  ~, 
\label{def:Ds}	
\ee
where the subscript $B$ stands for the binary defined by escaper $s$. This definition of the reference term will be motivated below.

\presub {\bf Properties of reference term}. The defined reference term has the following properties \begin{enumerate}
\item Physically motivated.
\item Regularizes $\sig$.
\item Proportional to $\sig_0$.
\end{enumerate}
These properties are discussed in the motivation section below.

\presub {\bf Alternative schemes and normalization}. The regularized value depends on the choice of the reference part, a choice known as the regularization scheme. The divergence of $\sig$ is proportional to $\sig_0$ \eqref{div_sig2}, which suggests to consider a class of minimal reference terms that share this property, namely \be
\div \( \sig_\rf \) \propto \sigma_0  ~.
\label{div_ref}
\ee Within this class, the phase-volume has a single counter-term, and all regularization schemes differ only by a (finite) multiple of $\sig_0$.

Given the form of the counter-term we define a normalized and dimensionless phase-volume as follows
\be
	\tsig (\{ \al_a\}) := \frac{1}{(2 \pi)^6\, \sig_0} \,  \frac{|E|^4}{(M_3/M)^{3/2}} \, \sig(E; \{ m_a, \al_a\})
	\label{def:tsig}
	\ee
and similarly for $\sig_\rf$. In particular, the normalized regularized phase volume is denoted by \be 
	\hsig \equiv \tilde{\bsig}
\label{def:hsig}
\ee
The division by $\sig_0$ is introduced so that under a change of regularization scheme within the minimal class, $\hsig$ changes only by an additive constant, and the dependence on $E, m_a$ is factored out according to \eqref{reduction}. Finally, the numerical prefactor was chosen somewhat arbitrarily.

\presub {\bf Motivation for definition of the reference term and a discussion of it}. We chose the integrand of $\sigma_{\rf}$ to be given by the effective system for a well-separated escaper $s$. The first motivation for this is physical. The regularized phase-volume for systems with a single degree of freedom is equivalent to the delay time, and its definition uses asymptotic (free) motion as reference,  see e.g. \cite{flux_based} Section 3.2. Led by this example, and given that during asymptotic motion (entrance and exit) the three-body system is approximated by one of the effective systems, these systems are chosen as reference. Secondly, the reference should be chosen such that it has the same rate of divergence as the bare quantity. Since the rate of divergence is approximated by the effective systems \eqref{div_sig1}, the chosen integrand is appropriate.

Let us proceed to motivate the choice of the integration domains $D_s$. 

\begin{figure}
\centering \noindent
\includegraphics[width=10cm]{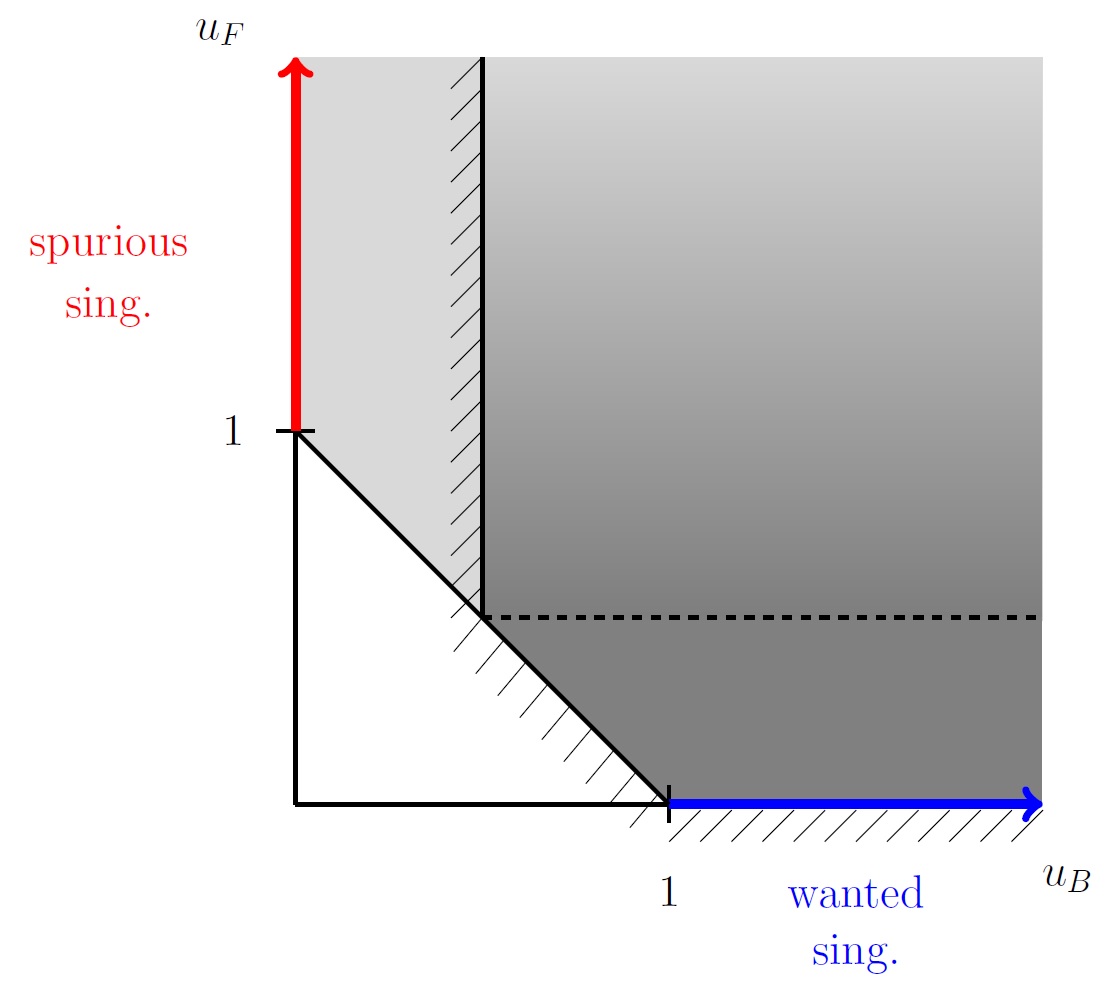} 
\caption[]{Various integration domains, $D_s \subset D'_s \subset D_{0,s}$, projected onto the $(u_B,u_F)$ plane defined in \eqref{def:u}. The grey-shaded region is $D_{0u}$, defined in \eqref{def:D0u}, and which is the projection of $D_{0,s}$, the maximal domain of definition for the integrand of $\sig_{\rf,s}$. It covers the wanted singularity, yet contains also a spurious singularity. The darker-grey region defined by requiring also $u_B \ge 1/2$ is the projection of $D'_s$. It avoids the spurious singularity, and regularizes $\sig$.  Finally, $D_s$,  our physically-motivated regularization scheme, is defined by further constraining to $E_B \le E/2$. It has the same projection as $D'_s$, but for $u_F > 1/2$ the unprojected $D_s$ is a proper subset of $D'_s$.}
\label{fig2:domain} \end{figure}

\presub {\it Singularities of the reference integrand}. First, we examine the integrand and its singularities. Let us denote by $\sig_{\rf,s}$ the $s$ summand in \eqref{def:sig_ref}.  The $\delta$-functions in \eqref{def:sig_ref} confine the integration domain to the so-called Hill region $D_{0,s}$ given by \be  
D_{0,s}(E) := \{ (\vec{r}_1, \vec{p}_1, \dots, \vec{r}_3, \vec{p}_3 ) ~ | ~ -V_F \equiv \frac{\alpha_B}{r_B}  + \frac{\alpha_F}{r_F} \ge -E\} 
\label{def:D0}
\ee
where by definition, the $B, F$ systems depend on $s$. Projecting phase space onto the $(u_B,u_F)$ plane, defined in \eqref{def:u} and shown in figure \ref{fig2:domain}, $D_0$ projects to into the $s$-independent domain
\be
D_{0u} := \{ (u_B,u_F) \in \IR^2 \, | \, u_B, u_F \ge 0, \, u_B+u_F \ge 1\} ~.
\label{def:D0u}
\ee

In the $(u_B,u_F)$ plane the reference term becomes \be
 \sig_{\rf,s} ( \{ \alpha_a \} ) = 16 \pi^2\,  \alpha_B^3\, \alpha_F^3\, \int_{D_{0u}}  \frac{du_F}{u_F^4}\, \frac{du_B}{u_B^4} \(u_B + u_F-1\)^2
 \label{sing_identification}
 \ee
 where we have carried the integrations over the momenta, as well as the directions of $\vec{r}_B, \vec{r}_F$, and presented the reduced form of the reference phase-volume, in analogy with (\ref{reduction},\ref{def:red_sig}). 
 
 This integral diverges at two loci, see figure \ref{fig2:domain}. The first is the line \be
 u_F=0 \qquad u_B \ge 1 \qquad \mbox{wanted} ~.
\label{wanted_div}
\ee
This represent configurations where the binary size is bounded, while the distance to the tertiary is very large. In these hierarchical configurations, the effective system approximates well the full system. This divergence is wanted because it cancels the divergence of $\sig_\br$, and we would like to keep it within the integration domain, as much as possible. 

The second divergence is along the line  \be
 u_B=0 \qquad u_F \ge 1 \qquad \mbox{spurious} ~.
\label{spurious_sing}
\ee
It is the image of the wanted divergence under the $B \leftrightarrow F$ symmetry of the effective system. It represents configurations where the binary center of mass is within a bounded distance from the tertiary, while the binary is very large. These are non-hierarchical configurations, where the effective system does not approximate the full system. In fact, $\sig_\br$ is not divergent there, and hence this divergence is unwanted, or spurious, and we would like to exclude if from the integration domain. 

Summarizing, through examination of  the integrand, we identified its maximal domain of definition, $D_{0,s}$, and its loci of divergence, the wanted and the spurious. 

\presub {\it Motivation for the choice of integration domain $D_s$}.  Recalling the motivation to subtract the asymptotic segments (entrance and exit) from the scattering time, we are led to impose $E_F \ge 0$, which describes scattering (hyperbolic relative motion).

However, such a reference term turns out to be insufficient to regularize $\sig$. The reason for this is that through subtraction of the asymptotic parts of the motion we express the average delay time. However, this quantity is known to diverge due to sub-escape excursions \cite{Agekian_Anosova_Orlov_1983}.

In order to regularize $\sig$, we must subtract also the phase-volume that describes sub-escape excursions. A second reason for that is that we started by wishing to approximate the regularized {\it chaotic} phase-volume, and hence we wish to remove these large regions of regular motion.

An excursion is hierarchical and hence we can continue to use $V_F$ for added part to $\sig_\rf$. In fact, an excursion is characterized by two conditions: a hierarchy of distances and a bound binary, namely $E_B<0$. The latter condition is required in order to distinguish from close encounters, which take place during a chaotic episodes, and in which a hierarchy of distances develops, but $E_B > 0$.    

In order to restrict the domain of integration of $\sig_\rf$ to hierarchical configurations, we should require $r_B \lesssim r_F$. In fact, it is convenient to remove complete orbits, and this can be achieved by refining the condition to involve the semi-major axes through $a_B \lesssim a_F$.

As mentioned above, it is suggestive to consider reference terms that belong to the minimal class \eqref{div_ref}. However, the last condition would lead to a reference term which does not belong to this class. This can be fixed, by changing the last condition into $-E_B \equiv \alpha_B/a_B \gtrsim \alpha_F/a_F \equiv -E_F$. In this way, the condition can be expressed in terms of the $u$ variables \eqref{def:u}, rather than the $r$ variables. At this point, we determine the precise form of the condition by replacing $\gtrsim \to >$. We could have inserted any constant of order 1, but this choice enjoys some extra symmetry (related to a reflection $u_B \leftrightarrow u_F$). When comparing with simulations, this reference implies a certain definition of the episodes which are identified at excursions.  

Altogether, we have \be
	E_B \le E_F
\ee 
Recalling that $E_B + E_F = E<0$, the condition has two other equivalent formulations \bse \begin{align}
 E_B &\le E/2 \label{EB_cond} \\
 E_F &\ge E/2 \label{EF_cond} ~.
\end{align} 
\ese
Formulation \eqref{EB_cond} immediately implies that $E_B <0$ and hence our condition fits both requirements for an excursion. The same formulation is the one that was quoted above in \eqref{def:Ds} for the definition of $D_s$. The preceding discussion demonstrates the physical motivation behind it (property 1 above). 

We can now return to examine the effect of this condition on the singularities of \eqref{sing_identification}. For $u_F \le 1/2$ the condition \eqref{EF_cond} becomes moot, and hence the wanted singularity is incorporated in full. On the other hand, \eqref{EB_cond} implies $u_B \ge 1/2$, and hence the spurious singularity is removed. This explains why this definition of the reference regularizes $\sig$ (property 2 above), and this would be confirmed by actual evaluations in the following sections. Property 3, $\sig_\rf \propto \sig_0$ is guaranteed by the construction, and will be seen to hold after \eqref{sig_ref_red}.

\subsection{Reduction of the reference}\label{refred}

In parallel to the reduction of $\sig$, we proceed to reduce $\sig_\rf$ through integration over momenta and the separation of the dependence on $E$. We find that $\tsig_\rf$, namely $\sig_\rf$ normalized according to \eqref{def:tsig}, is given by \be
 \tsig_\rf \( \{ \al_a\} \) = \frac{1}{2 \pi^4\, \sig_0} \sum_s \int d^3r_B\, d^3r_F \int_{\eps \le -1/2} d\eps \,  \sqrt{\frac{\alpha_{B,s}}{r_B}+\eps}^+ \sqrt{\frac{\alpha_{F,s}}{r_F}-1-\eps}^+ ~.
 \label{sig_ref_red}
\ee
In order to reach this, one expresses the phase space measure in terms of $\vec{r}_B,\,  \vec{r}_F,\, \vec{p}_B$ and  $\vec{p}_F$ through $\( \prod_{a=1}^3\, d^3 r_a \, d^3 p_a \) \delta^{(3)}(\vec{R}_{CM}) \, \delta^{(3)}(\vec{P}_{CM}) = d^3 r_B\, d^3 r_F \, d^3 p_B\, d^3 p_F$. Next, one inserts partitions of unity in the form $1=\int dE_B \,  \delta(H_B-E_B)  =\int dE_F \, \delta(H_F-E_F)$, where $H_B := p_B^2/(2 \mu_B) - \alpha_B/r_B$ and 	$H_F := p_F^2/(2\mu_F) - \alpha_F/r_F$.  We define $\eps:=E_B/|E|$. Finally, the integrations over $d^3p_B$ and $d^3p_F$ are performed. 

Changing the integration variables into the $u$ variables \eqref{def:u}, leads to $\tsig_\rf \( \{ \al_a\} \)$ being independent of $\al_a$ and hence $\sig_\rf \propto \sig_0$, as anticipated.

\presub {\it Definition of $\sig_\rfp$}. We find it convenient to implement the regularization through a subtraction of a different reference, denoted $\sig_{\rfp}$, accompanied by a compensation term $\Delta \sig$, namely \be
 \bsig \equiv \sig - \sig_\rf = \( \sig - \sig_{\rfp} \mbox{ at integrand level } \) + \Delta \sig 
 \label{ref_prime_method} 
  \ee
where \be
\Delta \sig :=  \sig_{\rfp} - \sig_\rf \mbox{ at integrand level }.
\label{def:Delta-sig}
\ee

We define $\sig_{\rfp}$  by integrating \eqref{def:sig_ref} over a domain $D'_s$ defined by \be
 D'_s := \{u_B \ge 1/2\} \subset D_0
 \label{def:Dp}
\ee

Just like $\sig_\rf$, $\sig_{\rfp}$ regularizes $\sig$ and is proportional to $\sig_0$. The benefit is that the $\eps$ integration can be performed analytically, yielding the following reduced form \be
 \tsig_\rfp \( \{ \al_a\} \)  = \frac{1}{16 \pi^3\, \sig_0} \sum_s \int_{\al_{B,s}/r_B \ge 1/2} d^3r_B\, d^3r_F \( \frac{\al_{B,s}}{r_B} + \frac{\al_{F,s}}{r_F} - 1\)_+^2 ~.
\label{sig_refp_red}
\ee

The compensator $\Delta \sig$ is independent of $\al_a$ and is given by \bea
	\Delta \hsig &=& \frac{8}{\pi^2} \int_{1/2}^\infty du_B \int_{1/2}^\infty du_F \int_{-1/2}^{u_F-1} d\epsilon \, \frac{\sqrt{u_B+\epsilon} \,\sqrt{u_F-1-\epsilon}}{u_B^4 \, u_F^4} \non
		 &=& \frac{20}{9 \pi} = 0.707355 \dots 
\label{def:Delta-hsig}
\eea
The integral is reached through the $u$ variables and its evaluation is detailed in appendix \ref{app:delta_sig}.


\section{Evaluation}
\label{sec:eval}

In the previous section, we have defined our regularization scheme. Now, we proceed to evaluate it: first analytically, and then numerically.  

\subsection{Analytic integrations}
\label{ai}

In this subsection, we show that the integration over several of the phase space variables can be performed analytically.

\presub {\bf Definition and momenta integration}. The bare phase-volume $\sig(E)$ was given in \eqref{def:sigE}, and the result of the integration over momenta was given in \eqref{momenta_integ}.

\vspace{0.5cm} The following steps are closely related to the natural dynamical reduction of \cite{DynRed}.

\noindent {\bf Planar Reduction}. The positions of the three bodies define a triangle, and hence a plane. It turns out to be possible to reduce the three dimensional coordinate integrals in \eqref{def:sigE} to the two dimensional integrals defined over this plane. The derivation is presented in the appendix \ref{pra}. Performing the planar reduction, we get \be 
			\int \( \prod_{a=1}^{3} d^3 r_a \) \, \delta^{(3)}(\vec R_{CM}) = \half \int \( \prod_{c=1}^{3} d^2 r_c \) \, \delta^{(2)}(\vec R_{CM}) ~ 2 \, A   \left(\int d\Omega \right) 
	\label{pr}
\ee 
where $A$ is the (positive) area of the triangle defined by the three bodies. In the RHS of \eqref{pr}, the integration is over 2d coordinates in the plane. $\int d\Omega$ denotes the angular integration over the the direction normal to the plane, $\hat{n}$. In this way, each three-body configuration is counted twice, at both $\hat{n}$  and $(-\hat{n})$ define the same plane. The $1/2$ factor accounts for that, so altogether $(1/2)\int d\Omega= 2 \pi$.  Inserting \eqref{pr} in \eqref{momenta_integ}, we get
\begin{equation}\label{ss}
	\sigma(E) = 16\pi^4 \left(\frac{M_3}{M}\right)^{3/2} \int \( \prod_{a=1}^3 d^2 r_a \) \, \delta^{(2)}(\vec R_{CM}) ~ A \,(E-V)_+^2 
\end{equation}

\presub {\bf The bi-complex position}. We introduce several changes of variables, going through the complex position vector, the bi-complex position and spherical coordinates thereof \cite{DynRed}. First, We introduce a change of variables to the complex position vector $\vec{w}$  and the planar center of mass coordinates as follows 
\begin{equation} \label{zv}
	\begin{split}
		\vec{w} &= \vec{r}_1 + e^{j\frac{2\pi}{3}}\, \vec{r}_2 + e^{-j\frac{2\pi}{3}} \, \vec{r}_3 \\		
		\vec{R}_{CM} &= \frac{1}{M}(m_1 \, \vec{r}_1 + m_2 \, \vec{r}_2 + m_3 \, \vec{r}_3 ) 
	\end{split}
\end{equation}
where $j^2  =-1$ is an imaginary unit (shortly, we shall introduce a second imaginary unit denoted by $i$). $\vec{w}$ decomposes into a pair of real vectors, the real and imaginary parts with respect to $j$, namely $\operatorname{Re} \vec{w}, \,  \operatorname{Im} \vec{w}$. This change of coordinates implies
\begin{equation}
	\prod_{a=1}^3 d^2 r_a = \frac{4}{3} ~d^2 \operatorname{Re} w ~ d^2 \operatorname{Im} w~ d^2 R_{CM}
\label{rw22}
\end{equation}

Next, we change the complex position vector $\vec{w}$ into the bi-complex position $w$ through \be
 w = \vec{w} \cdot (\hat{x} + i\, \hat{y} ) ~.
\ee
Now, $i^2 = j^2  =-1$ are two independent and commuting imaginary units (hence, such an algebra is known as bi-complex numbers). $w$ has four real components, i.e. the coefficients of $i, j, ij$, and the real part, and we denote these real components by $w_i, ~i=1,\dots,4$. Clearly \be
 	d^2 \operatorname{Re} w ~ d^2 \operatorname{Im} w = d^4 w ~.
\label{w224}
\ee

Lastly, we change to spherical coordinates on $\IC^2$, denoted by $r,\theta,\phi,\psi$, in the following way
\begin{equation}\label{ze}
	w = r~ e^{i\psi} \left[ \cos \frac{\theta}{2} ~e_R + e^{-i\phi} \sin \frac{\theta}{2} ~e_L \right]
\end{equation}
where, $e_{R,L} = \frac{\sqrt{3}}{2}(1\pm ij)$, and the ranges for the angles are $0 \le \theta \le \pi, \, 0 \le \phi \le 2 \pi, 0 \le \psi \le 2 \pi$ ($\psi$ was denoted $\psi_+$ in \cite{DynRed}). Below, we discuss the geometrical interpretation of these spherical coordinates. This coordinate change implies
\begin{equation}
	d^4w = \frac{9}{16}\, r^3  \sin\theta \, dr \, d\theta \, d\phi \, d\psi
\label{w_spherical}
\end{equation}

Implementing these coordinate changes in \eqref{ss} by substituting in (\ref{rw22},\ref{w224},\ref{w_spherical}) and performing the $\delta$-function integration over $\vec{R}_{CM}$, we find
\begin{equation}\label{siea}
	\sigma(E) = 12\pi^4 \left(\frac{M_3}{M}\right)^{3/2}\int r^3 \sin\theta ~A \, (E-V)_+^2 dr~ d\theta~ d\phi~ d\psi
\end{equation}

The triangle area and the potential $A, \,V$ are expressed in the spherical $r,\theta,\phi,\psi$ as follows
\begin{equation} \label{eaa}
	A  = \frac{\sqrt{3}}{4} \, r^2 \, |\cos\theta|
\end{equation}
and 
\begin{equation}\label{pea}
	V = -\frac{1}{r} \sum_{a=1}^{3}\frac{\alpha_a}{ \sqrt{1-\sin\theta\cos\left(\phi+\frac{2\pi(a-1)}{3}\right)}} 
\end{equation}
The derivation of these expressions is presented in appendix \ref{avd}. 

\presub {\bf Integrating over plane rotations and scaling}. 
The $\psi$ integration in \eqref{siea} is immediate: $\int d\psi = 2 \pi$. So
\begin{equation}
	\sigma(E) = 24\pi^5 \left(\frac{M_3}{M}\right)^{3/2} \int_0^\pi  d\theta \bar A \sin\theta \int_{0}^{2\pi}~ d\phi\int_0^{\bar V/E}dr~ r^5\left(E-\frac{\bar V}{r}\right)^2 
\end{equation}
where we display all $r$ dependence by defining $\bar{A}(\theta):=A/r^2$ and $\bar{V}(\theta,\phi):=r \, V$. Performing the $r$ integral, while assuming as always $E<0$, we get
\begin{equation}\label{avhe}
	\sigma(E) = \frac{2\pi^5}{5} \left(\frac{M_3}{M}\right)^{3/2}\frac{1}{|E|^4} \int_0^\pi  d\theta~ \bar A \sin\theta \int_{0}^{2\pi}~ d\phi~  \bar V^6
\end{equation}

We comment that the last equation can be interpreted as an integral of $\frac{1}{60} \, V^6$ over the projective configuration space. 

\presub {\bf Bare $\sig$ -- summary}. By performing a sequence of analytic integrations, we have reduced the expression for the bare phase-volume to an integral over the spherical coordinates $\theta, \phi$. Altogether, substituting  \eqref{eaa} and \eqref{pea} into \eqref{avhe}, and presenting the reduced $\sig \(\{ \al_a \} \)$ \eqref{reduction}, we get
\be
\label{bsia}
	\tsig\(\{ \al_a \} \) = \frac{\sqrt{3}}{640 \pi~ \sigma_0}
	 \int_0^{\pi/2}  d\theta \,  \sin 2\theta  \int_{0}^{2\pi} \, d\phi \,
	 \left(\sum_{a=1}^{3}\frac{\alpha_a}{\sqrt{1-\sin\theta\cos\left(\phi+\frac{2\pi(a-1)}{3}\right)}}\right)^6
\ee
In this expression, we have used the $\theta \to \pi - \theta$ reflection symmetry (corresponding to parity in configuration space) in order to restrict to the $0 \le \theta \le \pi/2$ hemisphere of the shape sphere.

We summarize the integration process in table \ref{table:integ_budget}. 

\begin{table}[ht]
\centering
\begin{tabular}{c|c|c}
		& No. of var.	 & No. of $\delta$-functions \\
\hline
Initial				& 18 & 7 \\
\hline \hline
Integ. Variables		& & \\
$\vec{p}_a$  		& 9	& 4 \\
$z_{cm}$, after planar trans. 	& 1 	& 1 \\
$d\Omega	$		& 2	&    \\
$\vec{R}_{cm}$, after $w$ trans.	& 2	& 2 \\
$\psi,\, r$			& 2	&  \\
\hline 
Total				& 16	&  7 \\
\hline \hline
Remaining		&  2	& 0  \\
\end{tabular} 
\caption{Budget of integration variables. The initial integral expression for $\sig$ \eqref{def:sigE} has 18 integration variables and 7 $\delta$-functions. Analytic integrations are possible for 16 variables in total, while integrating all $\delta$ -functions. It remains to integrate numerically over 2 variables: $\theta, \phi$.} 
\label{table:integ_budget} 
\end{table}

\presub {\bf Shape sphere}.  The $\theta,\phi$ coordinates parameterize a sphere known as the shape sphere, see \cite{Lemaitre_1952,Moeckel_Mont_2013,Montgomery2014,DynRed} and references therein. Every point on the shape sphere maps to a specific shape of the triangle defined by the three bodies (up to size). The overall length scale of the triangle is defined by coordinate $r$. The north and south poles describe right and left handed equilateral triangles. The equator describes the collinear configurations, as evident from the expression \eqref{eaa} for the area. Note that the expression \eqref{bsia} has singularities at three points on the $\theta=\pi/2$ equator at $\phi=0,\, 2 \pi/3, 4 \pi/3$.  These are the points where two of the bodies becomes coincident. We already described this in section \ref{bss}.
 
\begin{center}
	\begin{figure}[H]
		\includegraphics[scale=0.6]{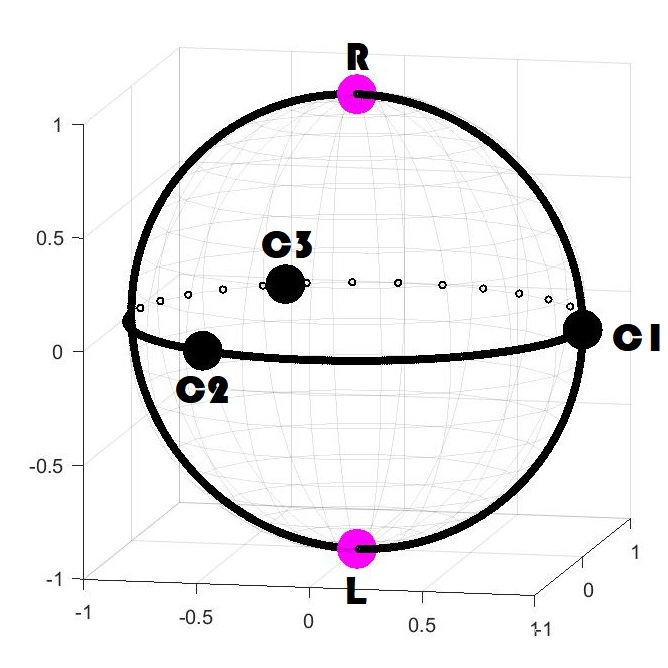}
		\caption{Shape sphere. Pink dots $R,L$ at poles describe right and left handed equilateral triangles, black dots $C1,C2,C3$ at the equator describe the three coincident configurations.}
		\label{shsp}
	\end{figure}
\end{center}

\presub {\bf Reference phase volume}.  For the $\rfp$ reference \eqref{def:Dp}, a procedure analogous to the one carried for bare $\sig$ produces the following reduction to an integral over the shape sphere \bea
 \tilde\sig_{\rfp,s} \( \{\al_a\} \) &=&  \frac{\sqrt{3} }{640\pi ~\sigma_0} \int_0^{\pi/2} d\theta \sin 2 \theta \int_0^{2 \pi} d\phi \non
  && \times \begin{cases}
 \bV_F^6 			& |\bV_F| \le 2 |\bV_B| \\
 (2 \bV_B)^4 \[ 10 (2 \bV_B)^2 -24 \bV_F (2 \bV_B) + 15 \bV_F^2 \] 	& {\rm otherwise}
\end{cases}
\label{sig_ref_integ}
\eea

\subsection{Numerical implementation and results}
\label{ni}

In this subsection, we describe the numerical integration for the evaluation of the regularized phase-volume.

It is required to numerically integrate over the shape sphere the difference of the integrands of $\tsig - \tsig_\rfp$, where $\tsig$ is given by \eqref{bsia} and $\tsig_\rfp$ is given by \eqref{sig_ref_integ}. Then we add the compensator $\Delta\hat\sigma$ to the result, as defined in \eqref{def:Delta-hsig}. The integrations are performed through the Numerical Integration (NIntegrate) command in Mathematica. The integrand is singular at 6 points around the equator of shape sphere, but the integral converges there and NIntegrate can handle that. The singular points consist of 3 coincidence points, which are the location of the singularity of $\sig_{\rm bare}$, and the 3 points of the spurious singularity, where $r_F \ll r_B$.

For a given mass set, say equal masses, we obtain finite results for the numerical integration, and after adding the compensator \eqref{def:Delta-hsig} according to \eqref{ref_prime_method}, we finally obtain a value for $\hsig$. 

\begin{center}
	\begin{figure}[H]
\includegraphics[width=.45\textwidth]{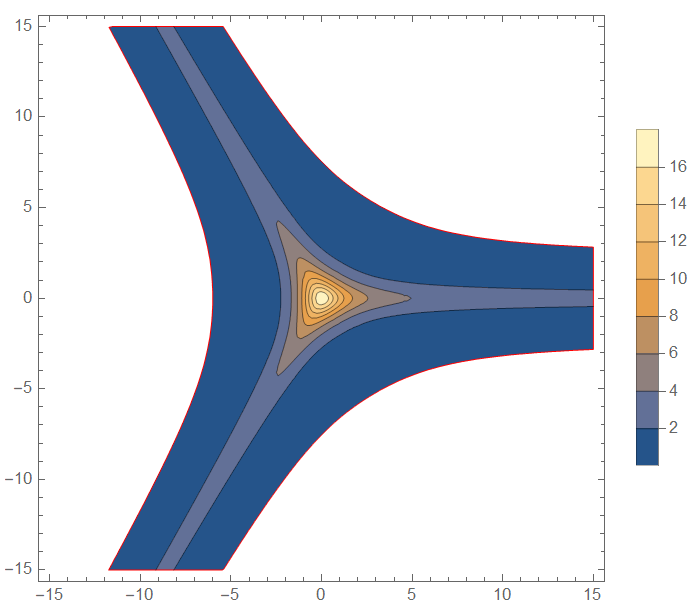} \hfill
\includegraphics[width=.45\textwidth]{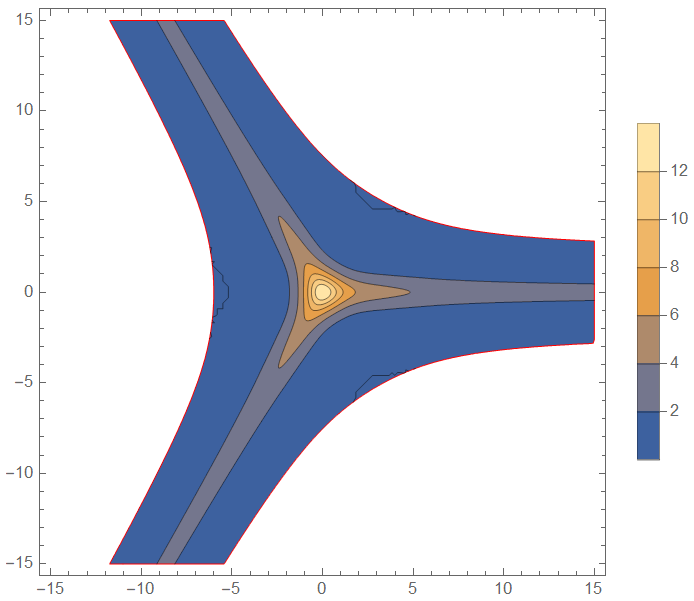} \\
\centering \includegraphics[width=.45 \textwidth]{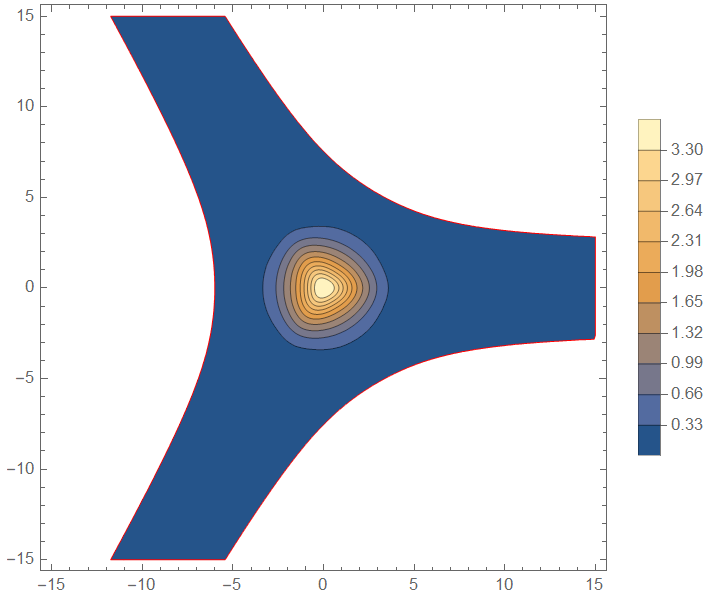}
		\caption{Illustration of the regularization of phase-volume. Shown are three contour plots of phase-volume density in a certain plane within configuration space. The shown densities are given by $(-V-1)_+^2$ \eqref{def:red_sig}, and so they do not contain certain measure factors. Top left: $\sig_\br \equiv (-V-1)_+^2$, top right: $\sig_\rf \equiv \sum_{s=1}^{3}(-V_{F,s}-1)_+^2$, bottom: $\bsig \equiv \sig_\br -\sig_\rf$.  Masses are equal, and units are such that $1=\al_a=|E|$. The shown plane is within geometry space, which describes the shape and size of the triangle formed by the three-bodies. It is parameterized by the spherical coordinates $r,\theta,\phi$ and the shown plane is specified by $1 = z \equiv r\,  \cos \theta$. 		
				The density of $\sig_\br$ is constant along the three pipes and that leads to a divergence of the integral. The same holds for $\sig_\rf$. Each pipe describes an asymptotic region, one such region for every possible escaper. In contradistinction, the density of $\bsig$ tends to $0$ along the pipes, and the integral over it converges, thereby providing regularization.}
		\label{fig:reg}
	\end{figure}
\end{center}

In order to gain some insight into the workings of the regularization, figure \ref{fig:reg} shows the densities of $\sig_\br$, $\sig_\rf$ and their difference.

\begin{figure}[H]
	\includegraphics[width=12cm]{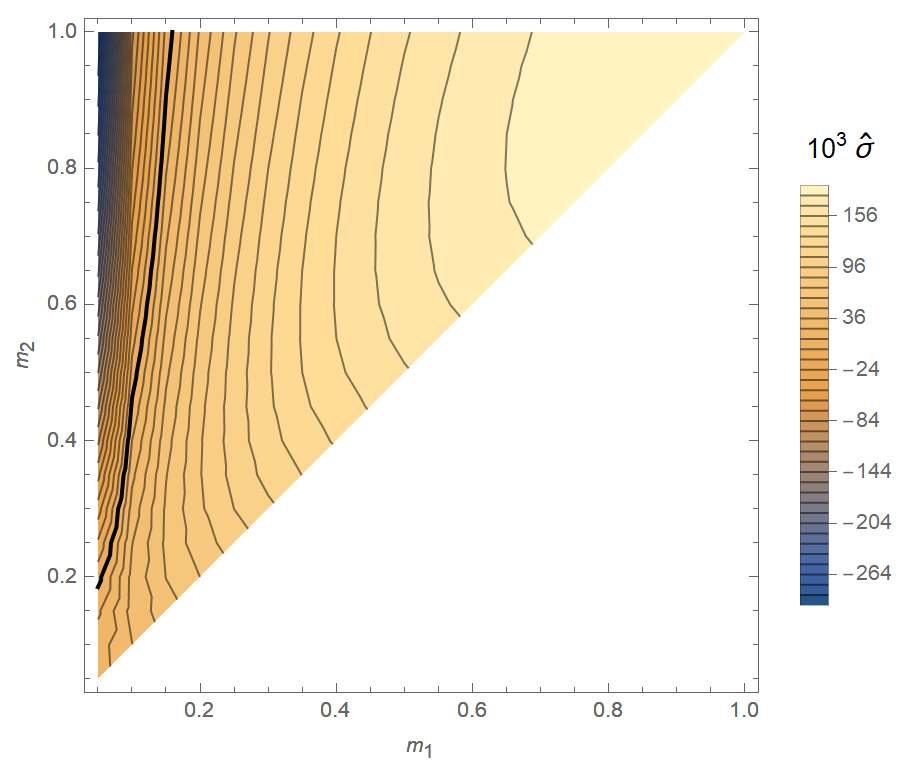} 
	\caption{Contour plot of the regularized phase volume of the 3d three-body problem with prescribed total energy, showing the normalized phase volume $\hsig$ \eqref{def:hsig} as a function of the masses $m_1 \le m_2 \le m_3=1$. The values depend on the choice of regularization scheme (within the minimal class defined by the property \eqref{div_ref}), but could change only by an overall additive constant.}
	\label{m1m2}
\end{figure}

Next, we evaluated $\hsig$ for numerous mass sets, and we present the results in the form of a contour plot in figure \ref{m1m2}. Without loss of generality, we consider mass sets within the triangular region $m_1 \leq m_2 \leq m_3 = 1$. We evaluated $\hsig$ over a uniform grid consisting of 210 points within this region, and used the values to generate the contour plot.

\presub {\bf Tests}. We validated the results by the following tests. \bi

\item Finiteness of the results, namely, regularization. This tests our regularization method, and in particular, the relative prefactor between the bare and reference terms.

\item Direct evaluation with $\tsig_\rf$. As a test, we evaluated $\hsig$ not only based on $\tsig_\rfp$, but also based on $\tsig_\rf$, and confirmed that we get the same values. This tests the change of variables into the shape sphere.

\ei

Let us explain a particularly important feature of the contour plot. We observe that in the test mass limit $m_1 \ll m_2, m_3$,  the regularized phase volume becomes negative. To illustrate this, we highlight the $\bar\sigma(E)=0$ contour line within the figure by making it bold. Literally, this means that the lifetime of the system vanishes. This is an indication that in such situations, the statistical approach to analyzing the three body problem ceases to be valid.


\section{Planar three-body}
\label{sec:planar}

In this section, we find the regularized phase-volume for the planar three-body system, by using the regularization scheme described for the 3d system. Planar three-body system is a special case where the initial velocities are along the plane within which the three bodies lie. The time evolution preserves the planar nature of the system. In this special case, much of the analysis performed for the general 3d system is analogous and somewhat simpler. So we choose to keep the section brief, and only present the main results and highlight distinct features. For more details, the reader can refer to the previous sections.

\presub {\bf Setup and Reduction}. Analogous to \eqref{def:sigE}, the phase-volume for the planar three body system (ignoring the total angular momentum) is defined as
\be
\sigma^{2d}(E)  := \int \( \prod_{a=1}^{3}\, d^2 r_a\, d^2p_a \)  \delta^{(2)}(\vec P_{CM}) ~ \delta^{(2)}(\vec R_{CM}) ~ \delta(H-E) ~.\label{bs2}
\ee
 Momentum integrations in the expression of $\sigma^{2d}(E)$ can be performed as
\bea 
\int \( \prod_{a=1}^{3}\,  d^2p_a \)  \delta^{(2)}(\vec P_{CM}) ~  \delta(H-E) &=&   \mbox{Vol}(\IS^3) \, \left(\frac{M_3}{M}\right) \, (2(E-V))_+ \non
&=&  4 \pi^2\, \left(\frac{M_3}{M}\right) \, (E-V)_+~.
\label{mi2}						
\eea
Factoring out the energy dependence in \eqref{bs2} we get
\be
\sigma^{2d}\(E;\{m_a, \al_a\}\) = 4 \pi^2\, \left(\frac{M_3}{M}\right) \, \frac{1}{|E|^3} \; \sigma^{2d}\( \{\al_a\} \) ~,
\label{red2}
\ee 
where $\sigma^{2d}\( \{\al_a\} \)$  is defined by \be
\sigma^{2d}\( \{\al_a\} \) :=  \int \(  \prod_{a=1}^{3} d^2 r_a \) ~\delta^{(2)}(\vec{R}_{CM}) \[ \( \sum_{a<b} \frac{\alpha_{ab}}{r_{ab}} \)- 1\]_+ ~.
\label{def:red_sig2d}
\ee

\presub {\bf Divergence}. 
The bare phase-volume \eqref{bs2} for the planar system diverges. Similar to the 3d system, the divergence originates from the hierarchical configurations. Switching to the effective variables $\vec r_B, \vec r_F$, the rate of divergence is given by
\be
\div \( \sigma^{2d}\( \{\alpha_a\} \) \) = \sum_{s=1}^3 \int^\infty d^2 r_{F,s} \int d^2 r_{B,s}  \( - V_{F,s}-1 \)_+ ~,
\label{div_sig1p}
\ee
where the effective potential $V_{F,s}$ was defined in \eqref{def:V_F}. By using similar arguments as for the 3d system, it can be easily seen that
\be
\div \( \sigma^{2d}\( \{\alpha_a\} \) \) \propto \sigma^{2d}_0 := \sum_{s=1}^3 \( \alpha_{B,s} \, \alpha_{F,s} \)^2 = (\alpha_1 + \alpha_2)^2 \alpha_3^2 + cyc. 
\label{div_sig2p}
\ee
Finally, switching to the $u_B,u_F$ variables defined in \eqref{def:u} we get \be
\div \( \sigma^{2d}\( \{\alpha_a\} \) \) = \sigma_0^{2d}  \int_0 \frac{d^2 u_F}{u_F^4}  \int \frac{d^2 u_B}{u_B^4}  \( u_B + u_F -1 \)_+ ~.
\label{div_sig3p}
\ee

\presub {\bf Regularization}. Analogous to \eqref{def:sig_ref}, we define the reference term for the planar system as \be
\sigma^{2d}_{\rf}(E) := \sum_{s=1}^3 \int_{D_s} \( \prod_{a=1}^{3}\, d^2 r_a\, d^2p_a \)  \delta^{(2)}(\vec P_{CM}) ~ \delta^{(2)}(\vec R_{CM}) ~ \delta(H_{F,s}-E) ~,
\label{def:sig_ref2d}
\ee
where the integration domains $D_s$ are defined to be all points in phase space such that \be 
E_B  \le E/2  ~.
\ee
This reference term has the following properties: it is physically motivated (see section \ref{regsec}), it regularizes phase-volume $\sigma^{2d}$, and it is proportional to $\sigma_0^{2d}$. For more details about the integration domains in the $u_B,u_F$ variables, the reader can refer to figure \ref{fig2:domain} and the associated discussion.  

We define the normalized and dimensionless phase-volume for the planar system
\be
\tsig^{2d} (\{ \al_a\}) := \frac{1}{(2 \pi)^4\, \sig_0^{2d}} \,  \frac{|E|^3}{(M_3/M)} \, \sig^{2d}(E; \{ m_a, \al_a\}) ~,
\label{def:tsig2}
\ee
and similarly define normalized and dimensionless $\tilde\sig^{2d}_\rf(\{ \al_a\})$. The normalized regularized phase-volume is denoted by $\hsig^{2d}$, which changes only by an additive constant under a change of regularization scheme within the minimal class defined by
\begin{equation}\label{mc2}
	\div \( \sig_\rf^{2d} \) \propto \sigma_0^{2d}~.
\end{equation}

\presub {\bf Reduction of the reference.} After switching to effective variables and then performing the momenta integrations, the normalized reference term reduces to 

\be
\tsig_\rf^{2d} \( \{ \al_a\} \) = \frac{1}{(2\pi)^2~\sig_0^{2d}} \sum_s \int d^2r_B\, d^2r_F \int_{\eps \le -1/2} d\eps \,  \Theta\left(\frac{\alpha_{B,s}}{r_B}+\eps\right) \Theta\left(\frac{\alpha_{F,s}}{r_F}-1-\eps\right) ~,
\label{sig_ref_red2}
\ee
where $\Theta$ is the Heavyside $\Theta$-function and $\eps:=E_B/|E|$.

For convenience, we implement the regularization by subtraction of a different but simpler reference term $\sig_\rfp^{2d}$, and account for this change by adding a compensator $\Delta\sigma^{2d}$. $\sig_\rfp^{2d}$ is member of the minimal class defined in \eqref{mc2}. For more details, the reader can refer to section \ref{refred}. 

The reference $\sig_\rfp^{2d}$ can be reduced to
\be
\tsig_\rfp^{2d} \( \{ \al_a\} \)  = \frac{1}{(2\pi)^2~\sig_0^{2d}} \sum_s \int_{\al_{B,s}/r_B \ge 1/2} d^2r_B\, d^2r_F \( \frac{\al_{B,s}}{r_B} + \frac{\al_{F,s}}{r_F} - 1\)_+ ~.
\label{sig_refp_red2}
\ee

And the compensator is found to be
\begin{equation}\label{do2}
	\begin{split}
			\Delta\hat\sigma^{2d} &=  \int_{1/2}^\infty du_F \int_{1/2}^\infty du_B \int_{-1/2}^{u_F-1} d\epsilon~  \frac{1}{u_B^3 u_F^3} \\ &=2
	\end{split} 
\end{equation}

\presub {\bf Analytic integrations.} We perform analytic integrations over several phase space variables for the bare phase-volume of the planar system. We find
\begin{equation}\label{bsi2}
	\tilde\sigma^{2d}\( \{\al_a\} \) = \frac{1}{16\pi ~\sigma_0^{2d}} \int_0^{\pi/2}  d\theta~ \sin\theta \int_{0}^{2\pi}~ d\phi~ \left(\sum_{a=1}^{3}\frac{\alpha_a}{ \sqrt{1-\sin\theta\cos\left(\phi+\frac{2\pi(a-1)}{3}\right)}}\right)^4
\end{equation} 
The expression of bare phase-volume for planar system has singularities at the same locations on the shape sphere (parameterized by $\theta,\phi$ coordinates) as the 3d system, refer to figure \ref{shsp} and the associated discussion. 

After performing the analytic integrations for the reference term $\sig_\rfp^{2d}$, we find
\bea
\tsig_{\rfp,s}^{2d}\( \{\al_a\} \) &=&  \frac{1}{16\pi ~\sigma_0^{2d}} \int_0^{\pi/2} d\theta \sin \theta \int_0^{2 \pi} d\phi \non
&& \times \begin{cases}
	\bV_F^4 			& |\bV_F| \le 2 |\bV_B| \\
	(2 \bV_B)^3 \[ -3 (2 \bV_B) + 4\bV_F \] 	& {\rm otherwise}
\end{cases}
\label{fr2}
\eea

\presub {\bf Numerical integration and Results.} We evaluate the regularized phase volume $\hat\sigma^{2d}$ by numerically integrating the difference between the integrands of $\tsig^{2d}$ and $\tsig_\rfp^{2d}$ over the shape sphere, as defined in \eqref{bsi2} and \eqref{fr2}. Then we add the compensator $\Delta\hat\sigma^{2d}$ to the result, as defined in \eqref{do2}. We use NIntegrate command in Mathematica for numerical integration. We generate contour plot of $\hat\sigma^{2d}$ using a uniform grid of 210 mass sets in the region $m_1\leq m_2\leq m_3=1$. 

\begin{figure}[H]
	\includegraphics[width=12cm]{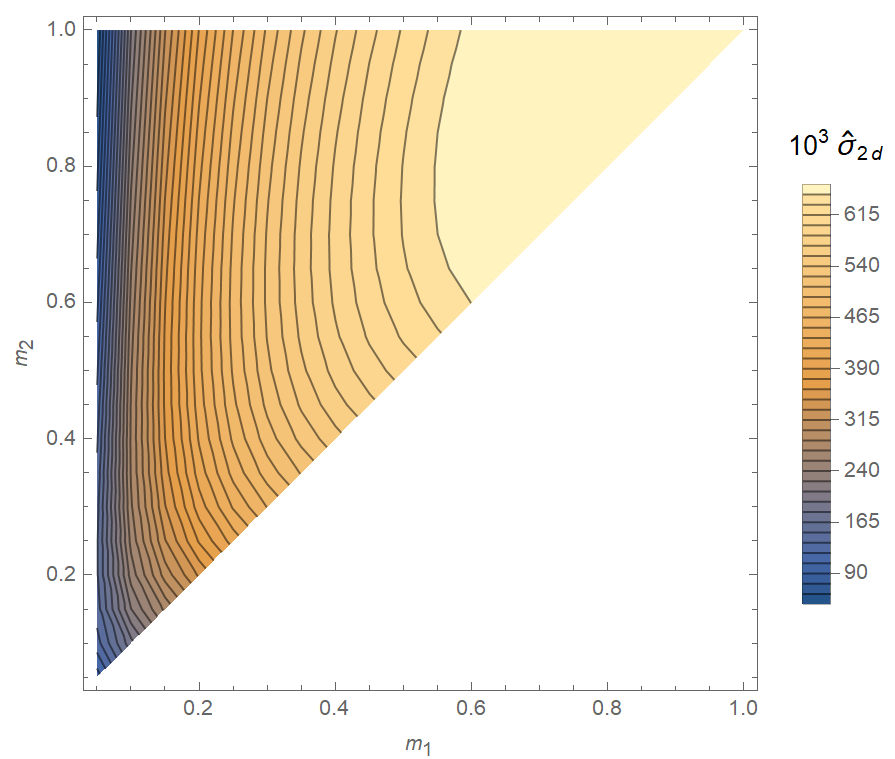} 
	\caption{Contour plot of the regularized phase volume of the \emph{planar} three-body problem with prescribed total energy, showing the normalized phase volume $\hsig^{2d}$ as a function of the masses $m_1 \le m_2 \le m_3=1$. The values are sensitive to the choice of regularization scheme (within the minimal class defined by the property \eqref{mc2}) only through an overall additive constant.}
	\label{m1m22}
\end{figure}

We observe that negative values of $\hsig^{2d}$ are not seen in the contour plot. However, the values do descend strongly as $m_1 \to 0$, and the current study cannot exclude that negative values do exist very close to the $m_1=0$ axis. 


\section{Discussion}
\label{sec:disc}

In this paper, we have fully defined a regularized phase-volume for the three-body problem $\bsig$. More specifically, we regularized $\sig(E)$, the phase-volume at given energy, by subtracting a reference term $\sig_\rf$ that is defined using the effective system of binary + tertiary (\ref{def:sig_ref},\ref{def:Ds}). We commented that it is possible to use alternative regularization schemes, but as long as they belong to the minimal class \eqref{div_ref} they will only shift $\bsig$ by a multiple of $\sig_0$, defined in \eqref{div_sig2}. 

Through analytic integrations, we have reduced the evaluation of $\bsig$ to an integral over the shape sphere (\ref{bsia},\ref{sig_ref_integ}). Finally, the latter integral was evaluated numerically. The finite result demonstrates the success of the regularization procedure. The regularization is illustrated in fig. \ref{fig:reg} which shows the densities of $\sig_\br$, $\sig_\rf$ and the difference. It can be seen how $\sig_\rf$ approximates  $\sig_\br$ over the asymptotic region, such that the difference is localized around the origin of configuration space.

$\bar\sigma$ was evaluated over all of the parameter space of possible mass sets, and presented as a contour plot in figure \ref{m1m2}, which is the main result of this paper. In section 4, the procedure was generalized to the planar three-body problem, and the contour plot of $\bsig$ as a function of the mass set is given in figure \ref{m1m22}.

We have performed strong tests for the correctness of our results -- see subsection \ref{ni}.

The positivity of $\bsig$ for comparable masses is not guaranteed, since the regularized value is a difference of two positive quantities. Rather, the obtained positivity is evidence for the consistency of the regularization and the flux-based theory. Moreover, it was found that $\bsig$ crosses into negative values in the test mass limit $m_1 \to 0$. Vanishing $\bsig$ implies diverging decay rates, and it is consistent with expectations that the non-hierarchical statistical theory ceases to be valid in the test-mass limit.

As far as we know, this is the first time that a regularized phase-volume of the Newtonian three-body problem is fully defined and determined. This has intrinsic interest, and is useful for normalizing the decay rates within the flux-based statistical theory. In addition, values of $\bar\sigma$ could be used to infer an averaged chaotic scattering time.

\presub {\it Open questions}. It would be interesting to evaluate $\sig(E,L)$ and to compare with decay rates measured through simulations. It would be interesting to obtain analytic expressions for $\bsig$ in the $m_1 \to 0$ limit.

\subsection*{Acknowledgments}

Part of this research was supported by the Israel Science Foundation (grant no. 1345/21). The work of S.M. was supported by the National Research Foundation of Korea grant NRF-2019R1A2C2084608. S.M. would like to thank the organizers of both `New Frontiers in Quantum Field Theory and String Theory' and `Advances in Theoretical Physics 2022' workshops for hospitality while this work was in progress.

\appendix
\newpage
\section{Compensator evaluation}
\label{app:delta_sig}

In this appendix, we detail the evaluation of the compensator \eqref{def:Delta-hsig}. In fact, we provide two derivations. 

The first way involves a reflection symmetry \bea
\Delta \hsig &=& \frac{8}{\pi^2} \int_{1/2}^\infty du_B \int_{1/2}^\infty du_F \int_{-1/2}^{u_F-1} d\epsilon \, \frac{\sqrt{u_B+\epsilon} \,\sqrt{u_F-1-\epsilon}}{u_B^4 \, u_F^4}   \non
	&=& \half  \cdot \frac{8}{\pi^2} \int_{1/2}^\infty du_B \int_{1/2}^\infty du_F \int_{-u_B}^{u_F-1} d\epsilon \, \frac{\sqrt{u_B+\epsilon} \,\sqrt{u_F-1-\epsilon}}{u_B^4 \, u_F^4} \non
	&=& \half  \cdot \frac{8}{\pi^2}  \cdot \frac{5 \pi}{9} = \frac{20}{9\pi}   ~.
\eea

The second way involves a direct integration with a convenient ordering of the three integrations \bea
	\Delta \hsig &=& \frac{8}{\pi^2} \int_{1/2}^\infty \frac{du_B}{u_B^4} \int_{-1/2}^\infty d\eps \, \sqrt{u_B+\eps} \int_{1+\eps}^\infty du_F \, \frac{\sqrt{u_F-1-\epsilon}}{u_F^4} \non
	&=& \frac{8}{\pi^2} \int_{1/2}^\infty \frac{du_B}{u_B^4} \int_{-1/2}^\infty d\eps \sqrt{u_B+\eps} \, \frac{\pi}{16} (1+\eps)^{-5/2} \non
	&=& \frac{8}{\pi^2} \int_{1/2}^\infty \frac{du_B}{u_B^4} \,  \frac{\pi}{24} \frac{(2 u_B-1)^{3/2}-1}{u_B-1} \non 
	&=& \frac{32}{3 \pi} \int_0^\infty du \, \frac{u (u^2+u+1)}{(u^2+1)^4(u+1)} = \frac{20}{9 \pi}
\eea
where in passing to the 4th line, we have changed variables according to $u^2=2 u_B-1$.


\section{Planar reduction}
\label{pra}

In this appendix, we derive \eqref{pr}. We start with the following rewriting of the LHS
\begin{equation} \label{s0}
	\begin{split}
		\( \prod_{c=1}^3 d^3 r_c \) \, \delta^{(3)}(\vec R_{CM}) &= \( \prod_{c=1}^3 d^2 r_c \) \, dz_1 \, dz_2 \, dz_3 \,  \delta^{(2)}(\vec R_{CM}) \, \delta\left(\frac{1}{M}(m_1 z_1 + m_2 z_2 + m_3 z_3)\right) \\
		&=\( \prod_{c=1}^3 d^2 r_c \) \, dz_1 \, dz_2 \, \delta^{(2)}(\vec R_{CM}) \left(\frac{M}{m_3}\right)~.
	\end{split}	
\end{equation}
Next, we express $dz_1$ and $dz_2$ in terms of rotations in the $x-y$ plane
\begin{equation}
	\begin{split}
		dz_1 &= d\vec\phi\times \vec r_1 \\ dz_2 &= d\vec\phi\times \vec r_2 ~,
	\end{split}	
\end{equation}
where $d\vec\phi$ is an infinitesimal vector in the $x-y$ plane. Since
\begin{equation}
	\begin{bmatrix}
		dz_1 \\ dz_2
	\end{bmatrix} = 
	\begin{bmatrix}
		y_1 & -x_1 \\
		y_2 & -x_2 
	\end{bmatrix}
	\begin{bmatrix}
		d\phi_x \\ d\phi_y
	\end{bmatrix} ~,
\end{equation}
we have
\begin{equation}\label{s1}
	dz_1 dz_2 = (x_1y_2-y_1x_2)d\phi_x d\phi_y ~.
\end{equation}

Area of the triangle formed by the three bodies is given by the determinant
\begin{equation}
	A= \frac{1}{2}\begin{vmatrix}
		x_1 & y_1 & 1\\
		x_2 & y_2 & 1\\
		x_3 & y_3 & 1
	\end{vmatrix}	~,
\end{equation}
which can be reduced using the following steps \footnote{We get the first equality by multiplying the third row by $\frac{m_3}{M}$, then adding to it $\frac{m_1}{M}$ times first row plus $\frac{m_2}{M}$ times second row.}
\begin{equation}\label{s2}
	\begin{split}
	    	A &= \frac{1}{2}\frac{M}{m_3}\begin{vmatrix}
			x_1 & y_1 & 1\\
			x_2 & y_2 & 1\\
			\frac{m_1 x_1 + m_2 x_2 + m_3 x_3}{M} & \frac{m_1 y_1 + m_2 y_2 + m_3 y_3}{M} & 1
		\end{vmatrix} \\ &= \frac{1}{2}\frac{M}{m_3}\begin{vmatrix}
			x_1 & y_1 & 1\\
			x_2 & y_2 & 1\\
			0 & 0 & 1
		\end{vmatrix}\\ &= \frac{1}{2}\frac{M}{m_3} (x_1 y_2-y_1 x_2)
	\end{split}
\end{equation}
Using \eqref{s1} and \eqref{s2} we get
\begin{equation}\label{s3}
	dz_1 dz_2 = 2A ~\frac{m_3}{M} ~d\phi_x d\phi_y
\end{equation}
Substituting \eqref{s3} in \eqref{s0} we get
\begin{equation}
	\begin{split}
			\( \prod_{c=1}^3 d^3 r_c \) \, \delta^{(3)}(\vec R_{CM}) &= 2A \( \prod_{c=1}^3 d^2 r_c \) \,  \delta^{(2)}(\vec R_{CM})~ d\phi_x d\phi_y\\ &= 2A \( \prod_{c=1}^3 d^2 r_c \) \, \delta^{(2)}(\vec R_{CM})~ d\Omega
	\end{split}
\end{equation}
where we rewrite $d\phi_xd\phi_y$ as $d\Omega$, which is the differential solid angle on the unit sphere. We have thus derived \eqref{pr}.

\section{Expressions for potential and triangle area}
\label{avd}

In this appendix, we derive the expressions for $A,V$ in terms of $r,\theta,\phi,\psi$ coordinates. For this, we need the expressions for the relative distances between the three bodies, i.e. $r_{12},r_{13},r_{23}$. First, we express them in terms of $w$ variables defined in \eqref{zv}
\begin{equation}
	r_{12} = ~\vline \frac{w-\eta \bar w}{1-\eta} \vline \quad ,\quad r_{13} = ~\vline \frac{w-\bar\eta \bar w}{1-\bar\eta} \vline \quad,\quad r_{23} = ~\vline \frac{w- \bar w}{\eta-\bar\eta}\vline
\end{equation}
where we use the following notation: $\eta=e^{j\frac{2\pi}{3}},~\bar\eta=e^{-j\frac{2\pi}{3}},~\bar w=w/.\{j\rightarrow-j\}$ and the absolute values are calculated with respect to $i$. Using \eqref{ze}, we rewrite in terms of $r,\theta,\phi,\psi$ coordinates
\begin{equation}\label{dc}
	\begin{split}
		r_{12} &= r \sqrt{1-\sin\theta\cos\left(\phi+\frac{2\pi}{3}\right)} \quad , \quad r_{13} = r \sqrt{1-\sin\theta\cos\left(\phi-\frac{2\pi}{3}\right)}\\ r_{23} &= r \sqrt{1-\sin\theta\cos\phi}
	\end{split}	
\end{equation}
From \eqref{dc}, it is straightforward to derive expression for $V$ in \eqref{pea}. 

To derive the expression for $A$ in \eqref{eaa}, we use the formula
\begin{equation}
	A = \frac{1}{4} \sqrt{(r_{12}^2+r_{13}^2+r_{23}^2)^2-2(r_{12}^4+r_{13}^4+r_{23}^4)}
\end{equation}
and substitute \eqref{dc} to derive \eqref{eaa}.

\bibliographystyle{unsrt}

\begin{thebibliography}{99}

\bibitem{Principia}
I.~Newton,  ``Philosophi{\ae}  Naturalis Principia
Mathematica'' (1687). 

\bibitem{Valt_Kart_book_2006}
M.~J.~Valtonen and H. Karttunen,
``The three-body problem,''
Cambridge University Press (2006).

\bibitem{Valtonen_etal_book_2016}
 M.~J.~Valtonen, J.~Anosova, K.~Kholshevnikov, A.~Myll\"ari, V.~Orlov and K.~Tanikawa,
``The Three-body Problem from Pythagoras to Hawking,''
Springer (2016).

\bibitem{Poincare_1890}
H.~Poincar\'e,
 ``Sur le probl\`eme de trois corps et les \'equations de la dynamique,''
 Acta Math.\, {\bf 13} (1890).

\bibitem{Agekyan_Anosova_1967}
T.~A.~Agekyan, Z.~P.~Anosova,
 ``A study of the dynamics of triple systems by means of statistical sampling,''
 Astron.\ Zh.\ {\bf 44} 1261 (1967).

\bibitem{Monaghan_1976} 
  J.~J.~Monaghan,
  ``Statistical-theory of the disruption of three-body systems - I. Low angular momentum,''
  Mon.\ Not.\ Roy.\ Astron.\ Soc.\ {\bf 176}, 63 (1976).

\bibitem{Stone_Leigh_2019} 
  N.~C.~Stone and N.~W.~C.~Leigh,
  ``A statistical solution to the chaotic, non-hierarchical three-body problem,''
  Nature {\bf 576}, no. 7787, 406 (2019).
  doi:10.1038/s41586-019-1833-8 .

\bibitem{Ginat_Perets_2020}
Y.~B.~Ginat and H.~B.~Perets,
 ``An Analytical, Statistical Approximate Solution for Dissipative and non-Dissipative Binary-Single Stellar Encounters,'' Phys.\ Rev.\ X, {\bf 11}, 031020 (2021)
 doi: 10.1103/PhysRevX.11.031020 .

\bibitem{flux_based}
B.~Kol,
``Flux-based statistical prediction of three-body outcomes,''
Celestial Mech.\ Dyn.\ Astron., Volume {\bf 133}, 17 (2021)
doi:10.1007/s10569-021-10015-x
[arXiv:2002.11496 [gr-qc]].

\bibitem{MTL_2020}
	V.~Manwadkar, A.~A.~Trani and N.~ W.~C.~Leigh,
``Chaos and L\'evy Flights in the Three-Body Problem,''
Mon.\ Not.\ Roy.\ Astron.\ Soc.\ {\bf 497}, 3694 (2020)
doi:  10.1093/mnras/staa1722 .

\bibitem{simulate}
V.~Manwadkar, B.~Kol, A.~A.~Trani and N.~W.~C.~Leigh,
``Testing the flux-based statistical prediction of the three-body problem,''
Mon.\ Not.\ Roy.\ Astron.\ Soc.\ {\bf 506}, 692 (2021)
doi:10.1093/mnras/stab1689 .

\bibitem{Agekian_Anosova_Orlov_1983}
 T. ~A. ~Agekyan, Zh. ~P. ~Anosova and V. ~V. ~Orlov, 
``Decay time of triple systems,''
Astrophys.\ {\bf 19}, 66 (1983). Translation of Astrofizika {\bf 19}, 111 (1983).
doi: 10.1007/BF01005813

\bibitem{DynRed}
B.~Kol,
``Natural dynamical reduction of the three-body problem,''
[arXiv:2107.12372 [astro-ph.EP]].

\bibitem{Lemaitre_1952}
G.~Lema\^itre,
 ``Coordonn\'ees sym\'etriques dans le probl\`eme des trois corps,''
 Ext.\ Bull.\ Acad.\ roy.\ Belg.\  {\bf 120}, 582 (1952).

\bibitem{Moeckel_Mont_2013}
R.~Moeckel and R.~Montgomery,
 ``Symmetric regularization, reduction and blow-up of the planar three-body problem,''
 Pac.\ J.\ Math.\ {\bf 262}, 129 (2013).

\bibitem{Montgomery2014}
R.~Montgomery,
``The three-body problem and the shape sphere,''
Am.\ Math.\ Month.\ {\bf 122}, 299 (2015).

\end{thebibliography}

\end{document}